# Anisotropic weakly localized transport in nitrogen-doped ultrananocrystalline diamond films

Kunjal V Shah, Dmitry Churochkin, Zivayi Chiguvare and Somnath Bhattacharyya*

Nano-scale transport physics laboratory, School of Physics and DST/NRF CoESM,

University of the Witwatersrand, Private Bag 3, WITS 2050, Johannesburg, South Africa

Based on magneto-resistance studies performed over a wide range of magnetic fields as well as temperatures, we establish that correction to conductivity in heavily nitrogen doped ultrananocrystalline diamond (UNCD) films is dominated by anisotropic weak localization (WL) in three dimensions associated with a propagative Fermi surface. Also, low temperature electrical conductivity can show weakly localized transport in 3D combined with the effect of electron-electron interactions in these materials, which is remarkably different from the conductivity in 2DWL or strong localization regime. The corresponding dephasing time of electronic wavefunctions in these systems described as $\sim T^{-p}$ with $p < 1$, follows a relatively weak temperature dependence compared to the generally expected nature for bulk dirty metals having $p \geq 1$. The temperature dependence of Hall (electron) mobility together with an enhanced electron density has been used to interpret the unusual magneto-transport features and show delocalized electronic transport in these n-type UNCD films, which can be described as low-dimensional superlattice structures.

___________________________________________________________________________

*corresponding author: Somnath.Bhattacharyya@wits.ac.za

**I. Introduction**

Ultrananocrystalline diamond (UNCD) films have evolved to be a unique form of nano-structured carbon (diamond) where the conductivity as well as microstructure of the material may be controlled by incorporating nitrogen in the grain boundary (GB) regions [1,2,3,4]. In these materials previously observed conductivity crossover from localized to delocalized regime has been interpreted as the interplay of dimensionality stimulated by the structural change, e.g. widening of the GB, creating structural anisotropy and most effectively by introducing high density of free carriers (by nitrogen doping) [5]. However, the specific origin of the conductivity rise by a large factor without changing the atomic concentration of nitrogen in the heavily doped films remains unclear [6]. Some reports claimed that the origin of high conductivity in this material is related to the formation of π bonded $sp^2$ clusters and additional mid-gap states, which makes transport characteristics of UNCD films similar to a low dimensional disordered metal (like graphitic carbon) instead of degenerate semiconductors [7]. On the other hand, following a non-degenerate semiconductor model the conductivity of UNCD films prepared with a wide range of $N_2$ concentrations was analyzed using a combination of hopping (strong localization) and activated (band) conduction process in three dimensions (3D) in different temperature regimes [8].

For heavily doped UNCD films (obtained from 10-20% $N_2$ in source gases), on the basis of evaluation of temperature dependences of conductivity as well as magneto-resistance (MR), contributions from weak localization (WL) and also from variable range hopping (VRH) transport (strong localization) to the net conductivity have been considered [4]. WL correction to conductivity in the GB regions is proposed as a dominant mechanism, especially at very low temperatures although the details of film microstructures in connection with mechanisms including 2D WL to 3D WL interplay and electron-electron (*e-e*) interaction effect in both dimensions have not been elucidated [5]. Also, high values of carrier concentration (density greater than conventional 2DEG system) and a moderately high value of electron mobility was reported, however, the origin of these high values in disordered carbon systems remains unclear [1]. To the best of our knowledge, there is no definitive proof of the formation of an impurity band yet exhibiting a high value of diffusive mobility (as in semiconductor heterostructures) within confined GB structure of UNCD films, which limits the potential electronic device applications of these materials with excellent mechanical properties. More importantly temperature dependence of characteristic time and associated scattering & electron dephasing mechanisms have not been presented explicitly in UNCD films, which also restrict the application of these films for fast electronic processes.

In this report, for the first time, we have been able to show evidence of 3D anisotropic and correlated electronic transport in *n*-UNCD films through conductivity and MR measurements at much lower temperatures (down to 0.3K) and at higher fields (up to 12 Tesla) than reported earlier [1-4]. This work also details Hall resistance measurements in these films not only to establish delocalized transport in these films but also to explain the unusual magneto-transport and a weak temperature variation of conductivity of the films in the 3DWL framework.

This article is structured in different subsections including an introduction (sec. I) followed by details of experimental techniques (sec. II). The results and discussion part of the work in sec. III(A) describes temperature dependent conductivity at zero field initially on the basis of (i) strongly localized as well as weakly localized conductivity (ii) in isotropic and 2D systems however focuses on (iii) the transport in 3D systems. Sec. III(B) discusses on MR study of UNCD films in (i) strongly localized as well as (ii) weakly localized regime and gives an in depth analysis based on (iii) 3D anisotropic WL conduction. Sec. III(c) deals with temperature dependence of Hall mobility and carrier concentration. Conclusions of all analyses and discussion are presented in sec. IV.

## II. Experimental

We mainly focus on the heavily-doped UNCD films, i.e. UNCD$_{20N}$, which were prepared with 20%N$_2$ by gas-phase deposition whose properties are compared with UNCD$_{10N}$ films, prepared by introducing 10%N$_2$ gas in the reaction chamber [6]. The UNCD films of average thickness about 500 nm were deposited on fused quartz substrates. For transport property measurements using the van der Pauw configuration a rectangular area of 0.5 cm$^2$ were contacted on the four corners of the rectangle [see Fig. 1(a), inset]. The temperature dependence of conductivity and MR measurements were performed in the temperature range 0.3 K - 300 K and at fields up to 12 T using a He$^3$ probe on a completely automated and cryogen-free cryostat (Cryogenic Ltd. UK). Hall resistance measurements were performed at different temperatures by applying the magnetic field, $B = \pm 1$ T ($B \perp$ to the plane of the film and supplied current in the samples), and measuring voltages for current, $I = \pm 1$ µA, across the diagonals of the samples. The van der Pauw formula was then solved numerically for each *I-V* pair, using the method of nested intervals, to get the sheet resistance, $R_\square(0,T)$. It is noted that the reciprocity theorem was satisfied within less than 3% error, so the configuration was consistent with the requirements and the values obtained for $R_\square(0,T)$ could be used for the calculation of Hall mobility ($\mu$) and sheet carrier density $n_{2D}$.

## III. Results and Discussion

### (A) Zero field and low temperature conductivity

A semi-logarithmic plot of zero-magnetic-field resistance, $R(0)$, vs. *T* data for UNCD$_{20N}$ and UNCD$_{10N}$ samples is shown in Fig. 1(a). The deviations of the semi-logarithmic plots of $R(0)$ vs. *T* curves of Fig.1(a) from straight lines confirm a different nature of electrical transport in UNCD$_{20N}$ films when compared to typical 2DWL observed in isotropic disordered systems [11], for which straight lines in this plot are expected, particularly at low temperature side [see thick lines in Fig. 1(a)]. One of the possibilities for the deviation arises from the $T^{1/2}$ dependence of $\sigma(T)$ in addition to the $T$ independent $\sigma_0$ term [see Fig. 1(a) thin red line]. However, for UNCD$_{10N}$ samples this plot follows a straight line [thin blue line in Fig. 1(a)]. For these samples $R(0,T)$ data could not be fitted in the intermediate to high temperature range through a single process, namely VRH conduction [sec. III A(ii), Fig 1(b)], or WL correction [IIIA(ii), Fig 1(c) and inset]. Figure 1(c), inset shows conductance ($G$) vs. ln$T$ curves for UNCD$_{20N}$ from which their conductivity characteristics namely WL corrections were inferred. At the lowest measured temperatures (0.3 K), a finite value of $\sigma$ is clearly revealed with a minimum conductivity of about 200 $\Omega^{-1}$ cm$^{-1}$ for UNCD$_{20N}$ films. The conductivity values and their temperature dependence, i.e. $\sigma(T)$, of the measured samples were found to be consistent with previous reports [5, 8] down to 1.6 K, however we extended the measurement temperature range down to 0.3 K in order to redefine the charge transport mechanisms in these materials. Fig. 1(c), inset also shows the saturation of conductance for UNCD$_{10N}$ films at low temperatures.

**(i)** *Strongly localized (hopping) conductivity*

The zero field resistance $R(0)$ in the strongly localized (hopping) regime is given by the VRH formula $R(0) = R_0 \exp[(T_0/T)^{1/\gamma}]$ where the pre-factor $R_0$ may depend slowly on $T$ and the exponent $\gamma$ is 4 (Mott's law) in 3D in the absence of Coulomb interactions, and 2 (Efros-Shklovskii's law) in the presence of Coulomb interactions [9, 10, 11]. We have plotted the zero field $R(0,T)$ data recorded over the whole temperature range as a function of $T^{-1/4}$ and $T^{-1/2}$ in Fig. 1(b), and inset respectively, for both UNCD$_{20N}$ and UNCD$_{10N}$ samples. We observe that $R(0,T)$ data follow neither Mott's law nor Efros-Shklovskii's law below 20 K for UNCD$_{20N}$ and below 60 K for UNCD$_{10N}$ samples. Deviation from VRH transport at low temperature is more pronounced for UNCD$_{20N}$ than for UNCD$_{10N}$ films, which means hopping is not the dominant mechanism that explains the charge carrier transport in this system particularly at low temperatures. This claim is also verified from MR study discussed in section IIIB.

**(ii)** *Weakly localized transport in isotropic disordered materials and in two dimensions*

Here we check the applicability of conventional models of (isotropic) disordered metals or semiconductors to explain conductivities of UNCD films [10-12]. For weakly disordered metallic system the temperature dependent conductivity can be expressed as a sum of WL correction (*L*) and *e-e* interaction (*I*) [9,10], which are additive to the Drude conductivity ($\sigma_0$) in the lowest order,

$$\sigma(0,T) = \sigma_0 + \Delta\sigma(0,T)_L^{2D/3D} + \Delta\sigma_I^{2D/3D}(0,T) \tag{1}$$

We attempt to make a clear distinction between 2DWL and 3DWL corrections and also the contributions from *e-e* interactions to $\sigma(T)$. The 2DWL fit in disordered metals is expressed in terms of the Thouless length $L_{Th}$ and a proportionality constant, *B'* as:

$$\Delta\sigma(0,T)_L^{2D} = \frac{e^2}{\pi h}\ln\left(\frac{1}{L_{Th}}\right) = B'\ln(T), \tag{2}$$

The 2D *e-e* interaction term can also be expressed as

$$\Delta\sigma(0,T)_I^{2D} = \left(\frac{e^2}{4\pi^2\hbar}\right)\left(2 - \frac{3}{2}F\right)\left[\ln\left(\frac{\tau k_B T}{\hbar}\right)\right]_{2D} \tag{3}$$

Here *F*, $\tau$ and $k_B$ represent electron screening factor in 2D, the relaxation time corresponding to *e-e* process and Boltzmann constant, respectively [11]. Since the thickness (*d*) of the samples is much greater than $L_{Th}$ it is difficult to observe 2D *e-e* interaction effect in the present system described in Eq. 3. To observe the signature of 2DWL in conductivity a *lnT* behavior from these samples is expected but this was not observed for UNCD$_{20N}$ films, particularly at the low temperature range [13, 14]. This non-linear behavior of *R*(0) vs. *lnT* curve, particularly in the low temperature regime can be explained by a 3DWL model combined with a significant effect of 3D *e-e* interaction using Eq. (4) to (7) [Fig. 1(c)].

For UNCD$_{10N}$ films, $R(0)$ follows a $lnT$ trend below 50 K, however an attempt to match corresponding the zero-field resistance, $R(0)$ vs. $T$ curves with $lnT$ behavior above 50 K, shows a deviation from linearity [Fig. 1(a)]. Most importantly, the non-linearity of $G(T)$ vs. $\ln T$ curves for UNCD$_{10N}$ samples shown in Fig. 1(c) inset clearly exhibits the inadequacy of this 2DWL correction to $\sigma(0,T)$ [explained by Eq. (2)], which is commonly applied in an isotropic system. In fact, UNCD$_{10N}$ samples behave very differently from ideal 2DWL description in the presence of magnetic fields, which will be discussed in sec. IIIB. Therefore, we do not perform any analysis involving quantum correction for UNCD$_{10N}$ samples.

**(iii) *Weakly localized transport in three dimension***

Now we would like to verify the transport mechanisms in 3D for UNCD$_{20N}$ films using Eq. (4) to (7). It should be noted that the value of the conversion pre-factor from 2D to 3D resistance appears to be very crucial in this analysis [11]. Namely, it is common that $R_\square(0,T)$ acts as the geometrical pre-factor for 2DWL expressions. We determined $R_\square(0,T)$ from measured data by using the 3D resistivity [$\rho^{3D}(0,T)$] divided by the sample thickness i.e. $R_\square(0,T)= \rho^{3D}(0,T)/d = R^{3D}(0,T) \times S/(d \cdot l)$, where $l$= sample length and $S$= cross-sectional area. Hence the geometrical pre-factor could be evaluated roughly for UNCD$_{20N}$ films as $R_\square(0,T) \approx$ 200 × (4.30 × 10$^{-11}$)/(1.09 × 10$^{-10}$) ≈ 400 Ω at the lowest measured temperature. The total 2DWL pre-factor is about 0.004, which is less than the value for conventional GaAs superlattice (SL) structures, consistent with the derived unusually large factor associated with conductivity correction. The value of $R_\square(0,T)$ could be even lower since $R^{3D}(0,T)$ decreases with increasing temperature, which further confirms that 2DWL is not a dominant transport mechanism in UNCD$_{20N}$ samples.

To verify the Coulomb (*e-e*) interaction effects in 2D and 3D first we estimated the characteristic length, $L_c = (hD/2\pi k_B T)^{0.5}$, for UNCD$_{20N}$ samples the diffusion constant $D \approx 1.6 \times 10^{-7}$ m$^2$ s$^{-1}$ and $L_c = (1.1 \times 10^{-9}) \times T^{-0.5}$ yielding a value of ≈ 1 nm at $T$= 1 K. For UNCD$_{10N}$ samples the diffusion constant $D \approx 5 \times 10^{-7}$ m$^2$ s$^{-1}$ and $L_c = (1.96 \times 10^{-9}) \times T^{-0.5}$, giving $L_c = 1.96$ nm at $T = 1$ K. In general, values of $L_c$ found in both samples are much smaller than $d$, which means that the e-e interaction effect in 3D expressed by Eq. (6), should also be considered. The value of $D$ obtained directly from Hall resistance measurements accounts for inelastic scattering (see in sec. IIIC). A fit to data for UNCD$_{20N}$ samples using 2DWL and 3D e-e interactions is shown in Fig. 1(c). To understand the contribution from e-e interactions i.e. [$\Delta\sigma^{3D}_I$] over $\Delta\sigma^{3D}_L(T) = e^2/\{\pi h L_{Th}\} = e^2/\{\pi h\} T^{p/2}$ in 3D systems we check for theoretical values of the exponent 'p' suggested previously as 3/2, 2 or 3 depending on the type of scattering, which correspond to e-e scattering in the dirty limit, clear limit and domination of electron-phonon scattering over inelastic scattering rate, respectively [9-11]. None of these values of $p$ could produce a proper fit to the present data [see Eq. (7)]. In fact, detailed MR analysis in sec. IIIB (iii) establishes the value of $p$ less than unity. To fit the data we use conductance expressed as:

$$G = G_0 + \left[\Delta\sigma(0,T)^{3D}_L + \Delta\sigma(0,T)^{3D}_I\right]\frac{S}{l} \qquad (4)$$

Here 
$$\Delta\sigma^{3D}_L = \frac{e^2}{2\pi^2\hbar}\frac{1}{L_\phi} \qquad (5)$$

and 
$$\Delta\sigma^{3D}_I = \left(\frac{e^2}{4\pi^2\hbar}\right)\frac{1.3}{\sqrt{2}}\left(\frac{4}{3} - \frac{3}{2}F\right)\sqrt{\frac{k_B T}{\hbar D}} = mT^{1/2} \qquad (6)$$

express corrections for isotropic 3DWL and 3D e-e interactions to conductivity, respectively [11]. Adding the discussed corrections to $\sigma_0$, in the disordered metallic regime the total conductivity [$\sigma(0,T)^{3D}_{Total}$] is then expressed as:

$$\sigma(0,T)^{3D}_{Total} = \sigma_0 + \left(\beta'T^{p/2}\right)^{3D} + \left(mT^{1/2}\right)^{3D} \qquad (7)$$

Temperature dependence of the conductance data for UNCD$_{20N}$ samples can be expressed as a sum of the term consists of temperature dependence of the dephasing length ($L_\phi$) [Eq. (5)] and 3D *e-e* interaction term [Eq. (6)] in isotropic 3DWL framework:

$$G(T) = G_0 + C_1\, T^{0.35} + C_2\, T^{0.5} \qquad (8)$$

The best fitting parameters are $G_0 = 4.3 \times 10^{-3}$, $C_1 = 5.2 \times 10^{-5}$ and $C_2 = 9.0 \times 10^{-4}$. The screening factor in Eq. 7, $F(\approx 0.3)$ determines the strength of screening potential, which is found to be lower than that of strongly correlated systems where *e-e* interactions provides a major contribution to $\sigma(T)$ [15,16,17]. From the best fit to data we derive $\mu(T) \sim (1.86 \times 10^{-3}) \times T^{-1}$ (m$^2$ V$^{-1}$s$^{-1}$) and the value of $D \approx 1.5 \times 10^{-7}$ m$^2$ s$^{-1}$, which is $T$ independent. Also the temperature dependence of $L_\varphi \sim 4.6 \times 10^{-8}$ T$^{-0.35}$, was found to be comparable with the estimation on the basis of MR measurements (see in sec. IIIB).

From this analysis we notice that the temperature dependence of $L_\varphi$ yields the value of the exponent $p$ ($\approx$ 0.7) less than unity, which cannot be explained by a 3D isotropic case. A very similar value of $p$ or $L_\phi$ can be found in the SL model as reported earlier where tunnel transport is predominant [13,14]. In that case we attempt to explain the conductivity data using a SL structure in the frame of 3D anisotropic propagative Fermi surface (PFS) model combined with the *e-e* interaction term [13,14]. This model was originally developed to explain transport in artificial SL having disorder and was recently applied to explain the unusual transport in nano-crystalline silicon films treated as a naturally formed SL [18]. A very similar kind of $\sigma(T)$ vs. $T$ dependence (and MR) has been found in the present UNCD system, which can be governed by tunnel transport. Therefore, to analyse the conductivity corrections over a wide range of temperatures we express the conductance as Eq. (8), which fits well the experimental data [Fig. 1(c)]. The fitting expression corresponding to PFS model, is similar to Eq. (4), however the second term includes an anisotropic coefficient and the parallel component of $D$, i.e., $D_{//}$ as described by the PFS model [13]. The first term also includes $G_0$ as well as the constant contribution from WL term described in the PFS model [sec.IIIB(iii)]. Alternatively, one can use diffusive Fermi surface (DFS) model, where the temperature variation of $L_\phi$ can yield a ln$T$ dependence of $\Delta\sigma(0,T)_I^{3D}$ [13]. However, the exact transport mechanism e.g., isotropic or anisotropic transport and applicability of PFS (or DFS) models in these materials cannot be interpreted based on $R$-$T$ data solely, but will also be done through the analysis of MR data given explicitly in sec. III(B). This further confirms the non-conventional $T$ dependence of $L_\phi$ i.e. ~$T^{-0.34}$ in the present system.

## (B) MR as a function of B

### (i) *Strong localization* (*Hopping transport*)

The novelty of negative MR [Fig.2 (a) - (c)] of the present UNCD films can be described by a nearly linear $B$ dependence, which is significantly different both from a $B^2$ or a ln$B$ dependence (not shown here). Further we observe that there is no upturn of MR into a positive region with the increase of the field and therefore, the model proposed by Bright [12,19], describing hopping interference and other interaction effects, is not truly applicable at high $B$ regions. Figures 2(a) and inset show the plots of normalized resistance, $[R(B)-R(0)]/R(0)$ as a function of $B$ and $B^{1/2}$, respectively at different temperatures, from which the critical field $B^*$ between the perturbative $B^2$ and the $B^{1/2}$ regimes was obtained. MR as a function of $T$ for UNCD films measured at different $B$ showed a $T^{0.5}$ dependence, which also has a weak field dependence [Fig. 3(a) and inset]. The values of $L_\varphi$ have been calculated at the low field regime where magnetic length $L_B^* = L_\varphi$ [see Fig. 3(b)]. We found that the value of the exponent of temperature is greater than 0.25, particularly in the low $T$ regions i.e. $L_\varphi$~ $T^{-0.3}$ [Fig. 3(b), inset]. This analysis confirms the exclusion of Mott's VRH theory for the present analysis unlike claimed in earlier reports [3, 4]. In fact, the best way to verify the presence of hopping transport in a material is by determining the temperature dependence of the characteristic length i.e., $L_\phi$ as well as its magnitude. The value of $L_\phi$ obtained from the PFS fit of MR data, which falls in the range of 20 to 70 nm and its temperature dependence of $\sim T^{-0.3}$ discards any major contribution of hopping transport in these materials [20]. Overall, our interpretation of MR data is different from that found in previous reports by Mares *et al.* [3] and recently Choy *et al.* [4] based on 2DWL and strong localization (hopping) correction to conductivity, respectively. Also, our MR data showed less steep slopes at low temperatures than found in previous reports, which suggested a marked difference of present conduction mechanisms from strong localization [4]. In addition to that we observe anisotropic 3DWL in our sample, which is fundamentally different as compared to 2DWL correction to conductivity observed by Mares *et al.* [3]. We interpreted our data in terms of anisotropic 3DWL to get the best fit yielding a proper value of $L_\phi$ [Fig. 3(c)].

**(ii)** *Weak localization in two dimensions*

For low $B$ and high $T$ (and in the absence of spin orbit motion & magnetic impurities) the generally expressed in 2DWL model for the 2D MR (for $B \perp J$) using the digamma function ($\Psi$) as $\Delta R_\square^{2D}(B,T)/R_\square(0,T) = -\rho_\square(0,T)e^2/\pi h \ [\Psi\{1/2 + B_\varphi(T)/B\} - \Psi\{1/2 + B_m(T)/B\}]$. The relaxation time $\tau_m$ corresponds to an elastic collision process defined through the relation $\tau_m = m^*\mu/e$ and can be estimated from the field $B_m(T) = \hbar/(4|e|D\tau_m)$. The (tunnel) effective mass of electron is given by $m^* = 0.06$ times rest mass of electrons found in an artificial carbon SL system [21]. The diffusion constant can be defined by means of Einstein relation as $D = \mu k_B T/e$ where $D$ (for UNCD$_{20N}$~ $1.5 \times 10^{-7}$ m$^2$/s) is found to be almost $T$ independent, obtained from experimental data of Hall mobility $\mu(T)$. Since $B_m$ remains fairly constant with temperature change at $B<B_m$ and $\tau_m$ (~$1/B_m$) is very small (~$10^{-16}$ s), the temperature dependence of MR and the field related to inelastic scattering $B_\varphi(T)$ can be used to determine the dephasing time $\tau_\varphi(T)$. Also, the applied transverse magnetic field quenches the WL effect when it becomes greater than $B_\varphi = h/(8\pi e D \tau_\varphi)$ and the logarithmic increase of the resistance due to WL can be suppressed. In our case, one can easily see that 2DWL model failed to describe the observed behavior of MR except at very low fields for UNCD$_{10N}$ samples [Fig. 2(a)]. It is clear that 2DWL model (digamma function) does not work in these samples except at very low fields [Fig. 2(b)]. This has also been verified using low as well as high values of dimensional pre-factors $R_\square(0,T)$. Anyway, from the low $B$ fit we find the typical value of dephasing time $\tau_\varphi = L_\varphi^2/D$ in the range of 10 ns to 30 ns, which can give a good estimation of $L_\varphi \approx 100$ nm for UNCD$_{10N}$ films at the lowest measured temperature [Fig. 3(b)]. The temperature dependence of $\tau_\phi$ is deviating from $T^{-1}$ widely in UNCD$_{10N}$ films [see Fig. 3(c), inset]. However, to verify the fit to MR data at high $B$ and low $T$ more accurately we use $\Delta\sigma^{2D}(B,T) = \Delta\sigma^{2D}_L + \Delta\sigma^{2D}_I$ fit expressed as high and low $B/T$ asymptotics [Fig. 2(b)].

$$\Delta\sigma_I^{2D}/\sigma^{2D}(0,T) = -R^{3D}(0,T)(\frac{S}{d.l})\frac{e^2 F}{4\pi^2 \hbar} g_2(h), \quad g_2(h) = \begin{cases} \ln(\frac{h}{1.3}) \text{ for } \frac{B}{T} \gg 1 \\ 0.084 h^2 \text{ for } \frac{B}{T} \ll 1 \end{cases} \quad (9)$$

In this expression $h = \dfrac{g\mu_B B}{k_B T} \cong 1.34 \dfrac{B}{T}$ depends on Bohr Magneton ($\mu_B$) and Landé $g$ factor [11]. From this analysis we find that low $B/T$ asymptotes are valid at low $T$ only but the high $B/T$ asymptote does not work at all to explain the MR data. However, the characteristics of 2DWL logarithmic behavior for high fields were observed neither in UNCD$_{20N}$ nor in UNCD$_{10N}$ samples even by adding the contribution from 2D $e$-$e$ interaction (assuming $F = 0.3$). On the other hand, the $B^{0.5}$ high field behavior was established with fairly good precision in 3DWL framework [Fig. 2(a), inset and Fig. 4(a)]. We have verified that the 2DWL model described using the digamma function does not fit MR data at all for UNCD$_{20N}$ samples. MR vs. $lnB$ plot may make a straight line fit for UNCD$_{20N}$ films at very low fields, however it may not enable us to have a good estimation of $L_\phi$.

**(iii) *3D anisotropic approach***

The negative MR throughout the measurement range, together with the negative temperature coefficient of resistivity suggested that the WL theory could be one of the plausible transport mechanisms, which should demonstrate ln$B$ dependence at low $B$ and high $T$, however it does not explain MR data in the high field regime [3-5]. The apparent similarity between experimental data recorded at low $B$ and 2DWL fit arises from the same asymptotics for 3DWL as well as 2DWL expression [Eq. (10)]. In contrast, the $B^{0.5}$ dependence of MR as predicted by 3DWL theory at high $B$ and low $T$ can establish these materials as 3D systems. From Fig. 2(a), inset and 4(a) we see that as we lower the temperature, the range of the temperature increases, where $B^{0.5}$ fit is followed [see Fig. 3(a)]. This is a major observation in the present work, which was not claimed earlier [3-5]. Therefore, ultimately we analyze the high field MR data in parallel transport (considering parallel part of $R$ only i.e., $R_\parallel$) in 3DWL PFS regime given by:

$$\Delta\sigma_\parallel^{3D}/\sigma(0,T) = \alpha\left(R^{3D}(0,T)\frac{e^2}{2\pi^2\hbar}\times\left(\frac{eB}{\hbar}\right)^{1/2}\right)f_3\left[\frac{(\hbar/eB)}{4D_\parallel\tau_\phi}\right] \qquad (10)$$

In this equation the Kawabata function expressed by $f_3(x=\{(\hbar/eB)/4D_\parallel\tau_\phi\}$ was proposed previously as $f_3(x)=\sum_{N=0}^{\infty}2\{[N+1+x]^{1/2}-[N+x]^{1/2}\}-[N+1/2+x]^{-1/2}$ [$N\equiv$ Landau quantum number] to account for 3DWL in anisotropic cases [13]. The pre-factor $\alpha$ describes an anisotropic transport in 3D [13].

To discuss on the applicability of the PFS model we argue that negative MR and 3DWL can be observed due to anisotropic distribution of disorder (or free carriers) in the grain boundaries of the UNCD films, which can be treated as a disordered SL [13]. In our previous work we made an effort to fit MR (linear $B$ dependent) data using 2DWL model for UNCD films, which we found as significantly different from graphitic sp$^2$-C films. These analyses suggest a strong possibility for 3DWL instead of 2DWL in UNCD films [5]. We attempted to describe a SL model for UNCD films very schematically in our previous report [5]. Under this SL framework we attempt to use the PFS model to explain the MR of UNCD$_{20N}$ (and also UNCD$_{10N}$) samples, which suggests the presence of 3DWL trajectories across the barriers (created by nano-diamonds, for example) [5,13]. The present approach is different from the SL model in the DFS limit where one can consider both coherent and incoherent transport separately [16]. Explaining MR data using DFS fit requires an in-depth analysis, which will be presented elsewhere. In general transport in UNCD$_{20N}$ samples can be governed by 3D anisotropic approach leading to an anisotropic PFS model [Eq. (10)] where the effect of 2DWL is considered to be negligibly small and refine the value of $\tau_\varphi$ as indicated in Fig. 3(c). Fitting to MR data by using Eq. (10), without using the anisotropy parameter $\alpha$, leads to a strong deviation from experimental data points although the functional dependence on magnetic field was quite similar to that observed in experiments [Fig. 4b]. This means isotropic 3DWL approach is not appropriate in the present case even though the isotropic model fits the $G(T)$ vs. $T$ data in Fig. 1(c). From Ref. 13 and 14 we understand that the distinction between an isotropic and anisotropic approach can be realized in the presence of magnetic fields only. Anyway, in the framework of anisotropic 3D PFS ($\alpha = 3$ at low temperatures) we explain the $G(T)$ by using $G(T) = 4.23 \times 10^{-3} + 1.7 \times 10^{-4} T^{0.32} + 8.4 \times 10^{-4} T^{0.5}$, which improves not only the quality of the fit but also yields a better a value of $L_\varphi$ comparable to that one obtained from MR data where the terms corresponding to $G_0$, 3D WL and $e$-$e$ interaction are found to be slightly modified [Fig. 1(c)] [14].

We present the change of values of $\tau_\varphi$ derived from the fitted data, which establishes the characteristic time of these materials, which falls in the range $T^{-0.4}$ to $T^{-0.7}$. The corresponding $L_\varphi$ is plotted in Fig. 3(c) showing the temperature dependence in the range $T^{-0.2}$ to $T^{-0.35}$. This means that the dominating transport mechanism in the present system may not be truly mediated by disorder enhanced $e$-$e$ scattering as reported earlier in 2D systems [11]. The value of $L_\varphi$ can be greater than 65 nm at the lowest measured temperature.

We notice that the overall behavior of the curves is very different from the WL model, proposed for 2D turbostatic graphitic carbon for low $B$, where $\tau_\varphi$ was strictly $T^{-1}$ dependent [12, 13]. Also for UNCD films, $\tau_\varphi(T)$ is very different from that of previously reported amorphous carbon films, which has been fitted using a power law as $\tau_\varphi \sim T^{-1.16}$ [5]. These results are also different from multi-walled carbon nanotubes where the effect of the disorder induced $e$-$e$ interaction in low-dimensional carbon systems was claimed [22,23,24]. We believe that $e$-$e$ interaction effect in SL systems with 3DWL corrections needs further investigations.

We notice that PFS fit is valid for UNCD$_{10N}$ samples to some extent [Fig. 2(c)]. After adding the 2D $e$-$e$ interaction to 3DWL PFS fit the quality of the fit to data (using $\tau_\phi$ as the only fitting parameter) does not improve very significantly [not shown here]. Since the amount of disorder in UNCD$_{10N}$ samples seems to be much more than UNCD$_{20N}$ samples the DFS model, understood from the reduced conductivity of the later samples as well as from microstructure analysis [6], which deals with incoherent (and coherent) transport will be more appropriate to explain MR data for UNCD$_{10N}$ samples.

**(C) Analysis of Hall coefficient and Hall mobility**

In order to understand the weak $T$ dependence of $\sigma(T)$ data we calculated Hall mobility and carrier concentration of UNCD films whose temperature dependence was found to be difficult to establish firmly in previous reports [1]. Measurement of Hall coefficient ($R_H$) with temperature has been found very effective for separating the contribution from $e$-$e$ interaction to WL in disordered metals [9, 10]. The room temperature value of $\mu \geq 1$ $cm^2(Vs)^{-1}$, and 3 $cm^2(Vs)^{-1}$ were obtained for UNCD$_{10N}$ and UNCD$_{20N}$ samples, respectively. 2D carrier concentrations for these samples are $n_{2D} \geq 10^{16}$ $cm^{-2}$ and $10^{15}$ $cm^{-2}$, respectively (Fig. 5). The present values of $\mu$, being up to 6 orders of magnitude greater than other weakly conducting a-C and undoped diamond-like carbon films can explain the observed weakly localized conduction overcoming strong localized regime [12]. The estimated values of $n_{3D} > 10^{19}$ cm$^{-3}$ for UNCD$_{20N}$ samples can explain the high conductivity of these films and suggests that these heavily doped UNCD films transport can reach the high end of the diffusive regime enabling band-like conduction. Although $n$ decreases as we lower the sample temperature, its minimum value remains greater than $1 \times 10^{13}$ cm$^{-2}$. This is a clear signature of delocalized conduction, which cannot be explained by the hopping conduction mechanism.

Due to absence of anomalous Hall resistance $R_H = (ne)^{-1}$ the negative values obtained confirm '$n$' type conduction in these materials. We confirm that the $T$ dependence of $n$ on $T$ is not strictly proportional to $T^{0.5}$ unlike the free electron model [16]. The proposed description is consistent with the model that applies to degenerate semiconductors as well as disordered metals [9, 10, 19]. Most importantly, for both films a nearly $T^{-1}$ dependence of $\mu$ has been observed, which produces temperature independent $D$ and was employed to fit MR data. The increase of $\mu(T)$ with lowering of temperatures for both samples is totally different from hopping conduction but resembles heavily doped semiconductor heterostructures of high density confined electron gas in 3D [13-15]. In the extreme case the behavior of these two parameters can be explained by weakly disordered metals [17, 19]. The overall $\mu$ vs. $n$ dependence can show a positive slope (can be extracted from Fig. 5) for the investigated samples. These results can be matched with the doped SL structures having intentional disorder which shows that the trend of $\mu(T)$ or $n(T)$ can appear not only due to doping effects but also structural changes in the samples [14]. We believe that by adding nitrogen preferentially during the growth of the samples an anisotropic micro-structure of doped-UNCD films can be formed. In previous reports on 2DWL claims, it was found very hard to realize the presence of 2D layer of electrons without establishing an SL structure [1-5]. In contrast to 2D WL orbits we attempted to show the presence of 3D WL orbits, which can be firmly confirmed by MR studies as presently reported.

## IV. Conclusions

Our observation is very consistent with previously reported experimental data [2-5]. Here we have presented a more generalized picture of temperature dependence of conductivity data, different from 2DWL model in isotropic metals but conforms to 3D or quasi-2D-like conductivity in artificial SL systems. Mobility analysis provides a self consistent MR picture. The effect of *e-e* interactions in conductivity is also understood in UNCD films, which possibly arises from the high density of free electrons in these materials. We believe this combination of high density (still lower than that of ordered metals) and particularly the enhanced mobility (greater than that of metals) can show true promises for diamond electronics. 3D nano-carbon super-structures as presently shown are widely different from bucky balls (0D), graphene (2D), nanotube networks or other exotic structures, from carbon family. It is very important for such structures to establish high coherent transport normal to the 2D planes and that requires low temperatures as well as high magnetic fields. Hence we establish the presence of 3D anisotropic nanostructures in a naturally grown carbon SL system, which is yet to be established in other carbon allotropes. UNCD films are superior to many other conventional semiconductors in terms of mechanical and tribological properties [25]. SL structures with anisotropic (metallic) transport in UNCD films can show potential applications in field emission devices, hetero-structures or fast quantum devices where the long dephasing time can play a crucial role. The observed mobility is sufficient for developing novel diamond based thin film transistors.

## V. Acknowledgements

We would like to thank to University Research Council (WITs) and DST/NRF (SA) nanotechnology flagship programme for financial assistance. SB is grateful to Late Alan Krauss, Argonne National Laboratory, for his inspiration and support to perform this work particularly towards the development of the superlattice model.

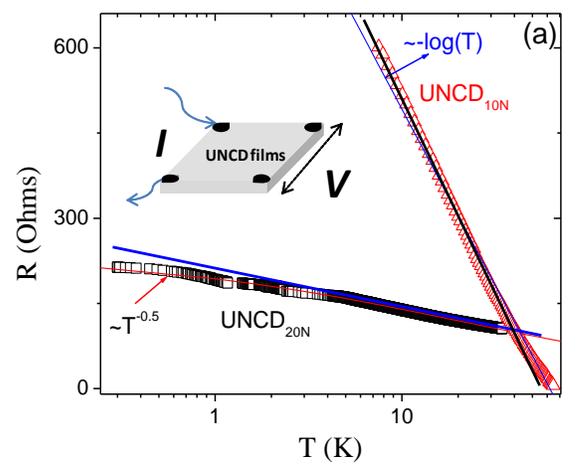

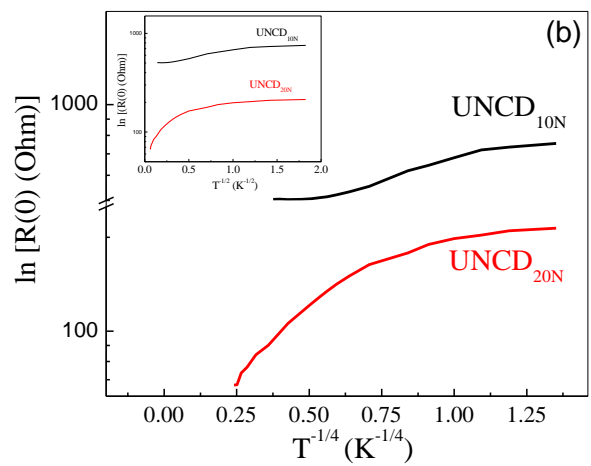

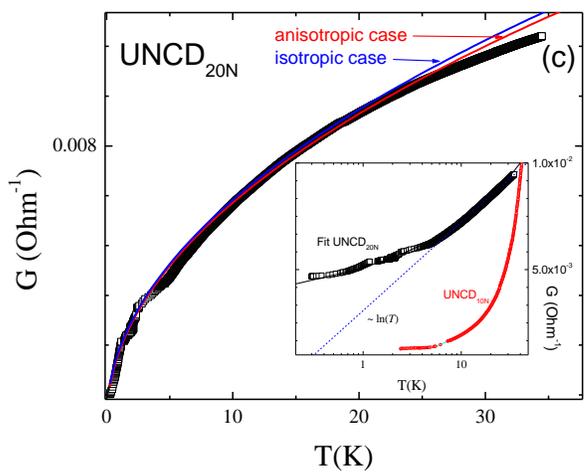

**Fig. 1(a), (b) and (c)**

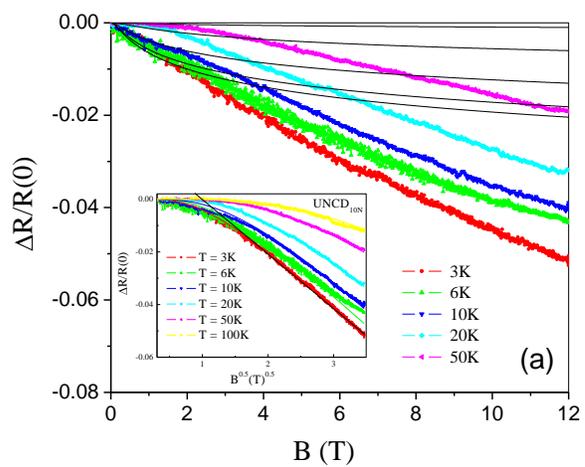

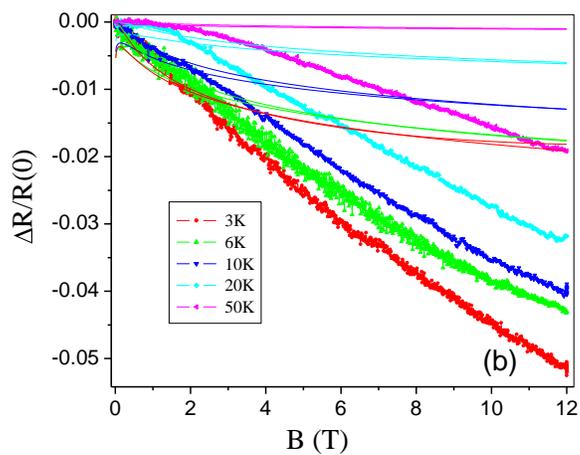

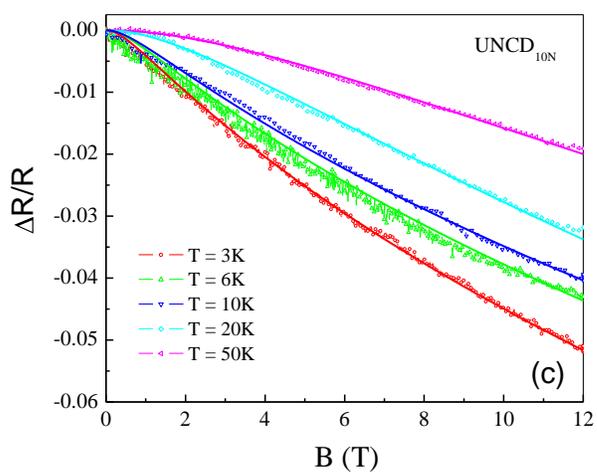

**Fig. 2(a), (b) and (c)**

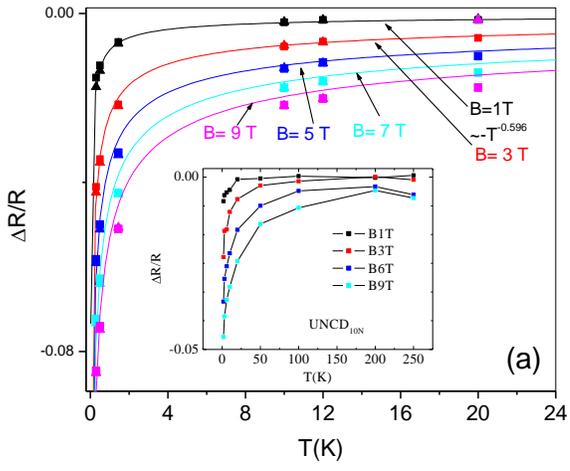
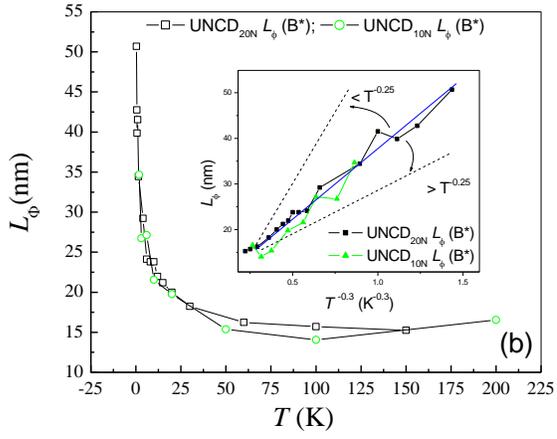
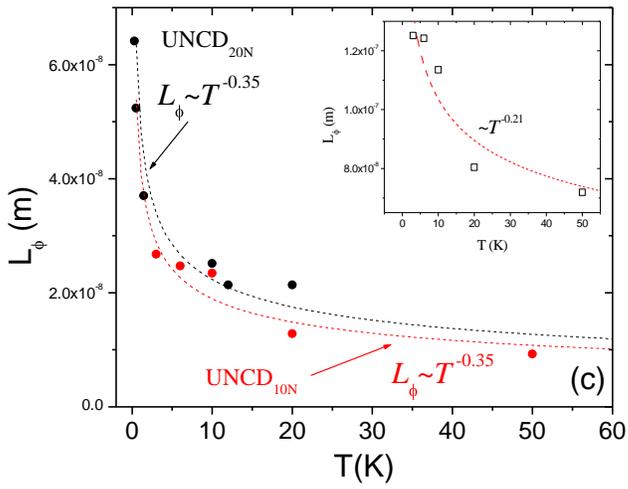

**Fig. 3(a), (b) and (c)**

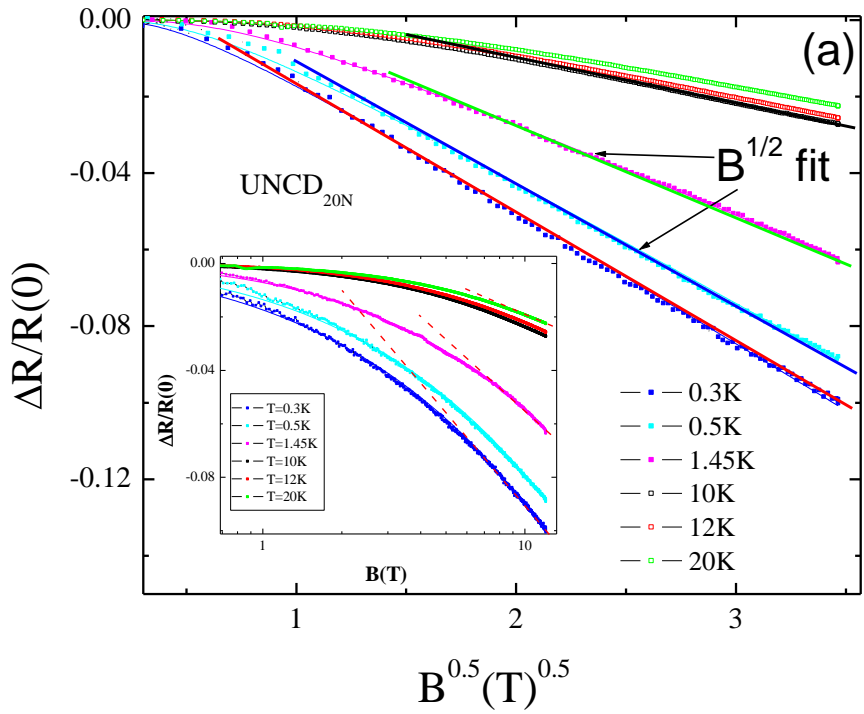
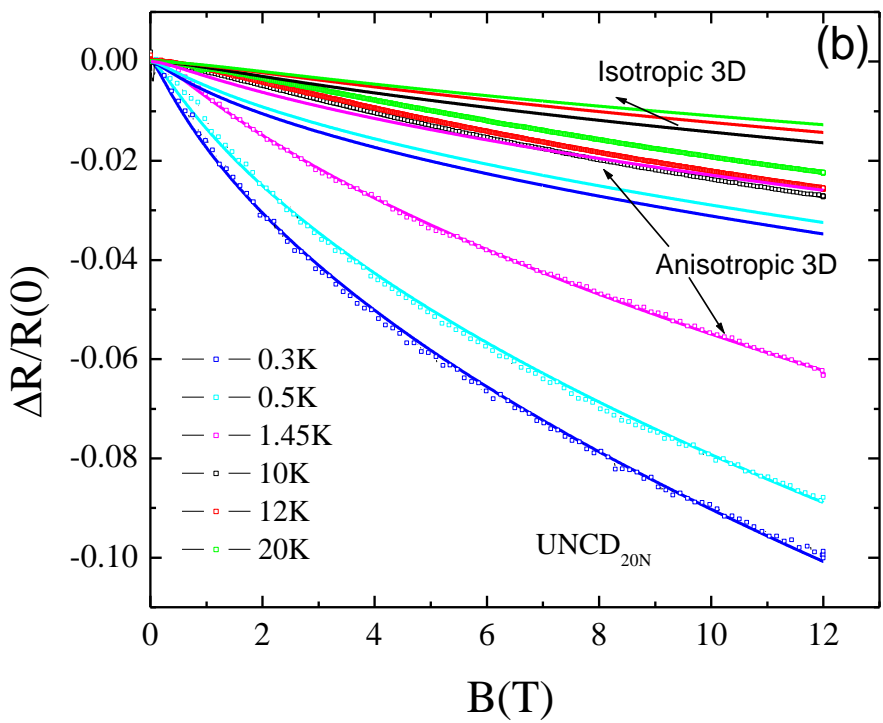

**Fig. 4 (a) and (b)**

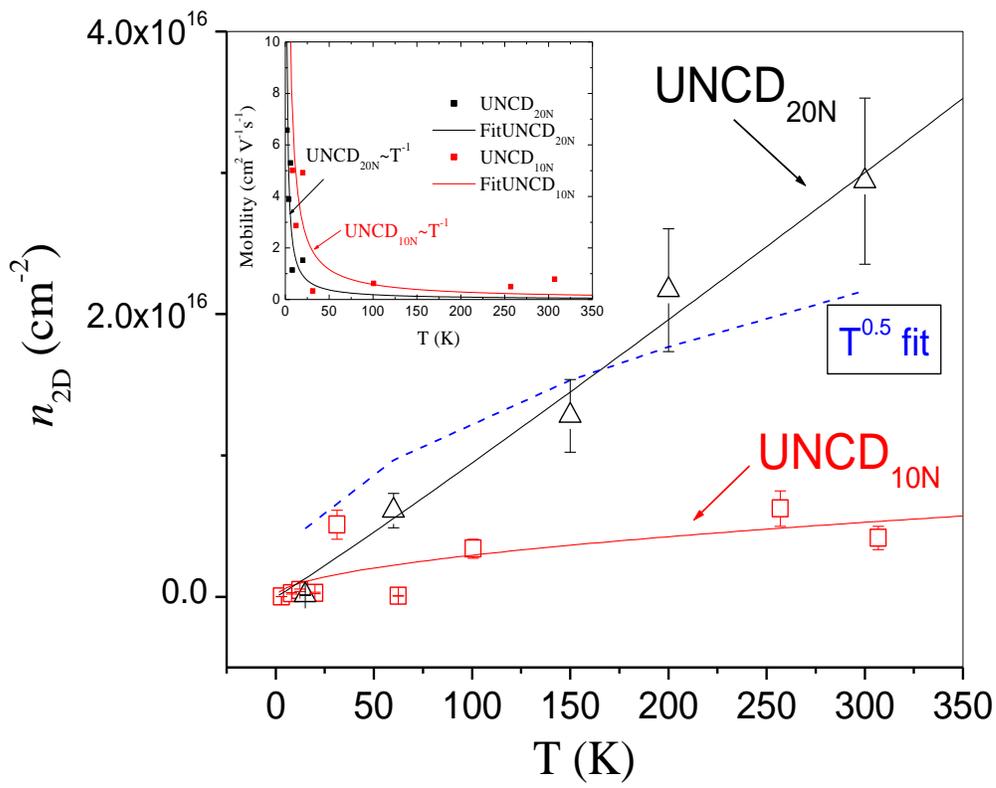

**Fig. 5**

**Figure Captions:**

**Fig. 1(a)** Resistance vs. temperature plot for UNCD$_{20N}$ and UNCD$_{10N}$ films and corresponding fit shows dominance of $T^{1/2}$ (to corresponding conductivity shown by red thin line) and $lnT$ nature (thin blue lines), respectively. Thick black lines are drawn to guide the eyes. **Inset:** Sample geometry used for the zero field resistance measurements.

**1(b)** 3D VRH fit for UNCD$_{20N}$ and UNCD$_{10N}$ samples at wide temperature range. **Inset:** 2D E-S fit to $ln(R(0))$ data.

**1(c)** Conductance vs. temperature plot of UNCD$_{20N}$ sample (black symbol) with anisotropic fit (red line) and isotropic fit (blue line) to data using 3DWL added with 3D $e$-$e$ interaction process [see text]. **Inset:** $log\ T$ scale shows a clear deviation of G from linearity (blue curve) at low temperatures for UNCD$_{20N}$ films and strong non-linearity for UNCD$_{10N}$ films (cyan curve).

**Fig. 2(a)** 2DWL (digamma) fit for UNCD$_{10N}$ samples with low value of dimensional pre-factor. **Inset:** In general the $B^{1/2}$ fits works well for UNCD$_{10N}$ films.

**2(b)** 2DWL plus 2D $e$-$e$ fit for UNCD$_{10N}$ films with different asymptotics. Solid lines are the high field asymptotics, which do not fit data properly. Dashed lines corresponding to low field asymptotics seem to fit the data.

**2(c)** 3D PFS fit (solid lines) for UNCD$_{10N}$ films, which shows that PFS model does not work fully in UNCD$_{10N}$ samples. Although 3DWL model fits the high $B$ regions it fails to work over the whole $B$ range.

**Fig. 3(a)** Temperature dependence of MR of UNCD$_{20N}$ samples recorded for different constant fields, which shows a $\sim T^{-0.6}$ behavior at low $B$. **Inset:** MR for UNCD$_{10N}$ samples shows a $\sim T^{-0.5}$ behavior at low $B$. MR recorded at different fields show no significant difference.

**3(b)** Dephasing length of electrons in UNCD$_{20N}$ samples calculated from critical magnetic fields ($B^*$). $L_\phi$ as a function of $T^{-0.3}$ dependence shows linearity.

**3(c)** Dephasing length of electrons in UNCD$_{10N}$ (red) and UNCD$_{20N}$ (black) samples calculated from 3DWL PFS fit. The temperature dependence of $L_\phi$ is given by $\sim T^{-0.37}$ for UNCD$_{10N}$ samples using 2DWL fit.

**Fig. 4(a)** $B^{0.5}$ high field asymptotic for UNCD$_{20N}$ films at low temperatures (solid lines). **Inset:** $log\ B$ scale does not produce any proper fit to data even at low temperatures.

**4(b)** Anisotropic 3D PFS WL fit for UNCD$_{20N}$ (data and solid lines). Dashed lines (in respective colors) show inapplicability of isotropic 3D PFS model to fit the data.

**Fig. 5** Sheet carrier concentration (symbols and solid thin lines) shows an increase with temperature. Dashed line shows a $T^{0.5}$ fit. **Inset:** Mobility shows an increase with lowering the temperature described as $\sim T^{-1}$ fit.

# REFERENCES

1. A. Williams, S. Curat, J. E. Gerbi, D. M. Gruen, and R. B. Jackman, Appl. Phys. Lett. **85**, 1680 (2004).

2. P. Achatz, O.A. Williams, P. Bruno, D.M. Gruen, J.A. Garrido, and M. Stutzmann, Phys. Rev. B **74**, 155429 (2006).

3. J.J. Mares, P. Hubik, J. Kristofik, D. Kindl, M. Fanta, M. Nesladek, O. Williams and D.M. Gruen, Appl. Phys. Lett. **88**, 092107 (2006).

4. T.C. Choy, A.M. Stoneham, M. Ortuno, and A.M. Somoza, Appl. Phys. Lett. **92**, 012120 (2008).

5. S. Bhattacharyya, Phys. Rev. B **77**, 233407 (2008).

6. S. Bhattacharyya, O. Auciello, J. Birrell, J.A. Carlisle, L.A. Curtiss, A.N. Goyette, D.M. Gruen, A.R. Krauss, J. Schlueter, A. Sumant, and P. Zapol, Appl. Phys. Lett. **79**, 1441(2001).

7. K. L. Ma, J. X. Tang, Y. S. Zou, Q. Ye, W. J. Zhang, and S.T. Lee, Appl. Phys. Lett. **90**, 092105 (2007).

8. S. Bhattacharyya, Phys. Rev. B **70**, 125412 (2004).

9. B.L. Al'tshuler and A. G. Aronov, Sov. Phys. JETP **50**, 968 (1979); B. L. Al'tshuler and A.G. Aronov, in Electron-Electron Interactions in Disordered Systems, edited by A. L. Efros and M. Pollak (North-Holland, Amsterdam, 1985).

10. N. F. Mott and M. Kavesh, J. Phys. C **14**, L659; A. Lee and T. V. Ramakrishnan, Rev. Mod. Phys. **57**, 287 (1985); N. F. Mott, 'Metal-Insulator Transitions', Taylor and Fransis, London (1990).

11. G. Du, V.N. Prigodin, A. Burns, J. Joo, C.S. Wang, A.J. Epstein, Phys. Rev. B **58**, 4458 (1998).

12. A. Bright, Phys. Rev. B **20**, 5142 (1979); V. Bayot, L. Piraux, J.-P. Michenaud, J.-P. Issi, M. Lelaurain, and A. Moore, Phys. Rev. B **41**, 11770 (1990).

13. W. Szott, C. Jedrzejek, and W.P. Kirk, Phys. Rev. Lett. **63**, 1980 (1989); A. Cassam-Chenai and D. Mailly, Phys. Rev. B **52**, 1984 (1995).

14. Yu. A. Pusep, M.B. Ribeiro, H. Arakaki, C.A. de Souza, S. Malzer, and G.H. Döhler, Phys. Rev. B **71**, 035323 (2005); J. Chiguito, Yu. A. Pusep, G.M. Gusev, and A.I. Toropov, Phys. Rev. B **66**, 035323 (2002).

15  K. Seeger, Semiconductor Physics, Springer, Wien, (1973).

16  J. M. Ziman, Principles of the Theory of Solids, Cambridge Univ. Press, Cambridge, UK (1972).

17  J.S. Dugdale, The Electrical Properties of Disordered Metals, Cambridge University Press, Cambridge, UK (1995).

18  K. Zhang, and W. Z. Shen, Appl. Phys. Lett. **92**, 083101 (2008)

19  V. Bayot, L. Piraux, J.-P. Michenaud, and J. -P. Issi, Phys. Rev. B **40**, 3514 (1989); L. Piraux, V. Bayot, J. P. Issi, M.S. Dresselhaus, M. Endo and T. Nakajima, Phys. Rev. B **41**, 4961(1990); R.T.F van Schaijk, A. de. Visser, S. G. Ionov, V. A. Kulbachinskii and V. G. Kytin, Phys. Rev. B **57**, 8900 (1998).

20  F. Tremblay, M. Pepper, D. Ritchie, and D.C. Peacock, Phys. Rev. B **39**, 8059 (1989).

21  S. Bhattacharyya, S.J. Henley, E. Mendoza, L.G-Rojas, J. Allam and S.R.P. Silva, Nature Materials **5**, 19 (2006).

22  K. Takai, M. Oga, H. Sato, T. Enoki, Y. Ohki, A. Taomoto, K. Suenaga, and S. Iijima, Phys. Rev. B **67**, 214202 (2003).

23  H. R. Shea, R. Martel, and Ph. Avouris, Phys. Rev. Lett. **84**, 4441 (2000).

24  N. Kang, J.S. Hu, W.J. Kong, L. Lu, D.L. Zhang, Z.W. Pan, and S.S. Xie, Phys. Rev. B **66**, 241403 (2002).

25  A.V. Sumant, D.S. Grierson, J.E. Gerbi, J.A. Carlisle, O. Auciello, and R.W. Carpick, Phys. Rev. B **76**, 235429 (2007); O.A. Williams, M. Nesladek, M. Daenen, S. Michaelson, A. Hoffman, E. Osawa, K. Haenen, and R.B. Jackman, Diamond and Relat. Mater. **17**, 1080 (2010).